\documentclass[runningheads]{llncs}
\usepackage[T1]{fontenc}
\usepackage{caption}

\usepackage{graphicx}
\usepackage{diagbox}
\usepackage{epstopdf}

\usepackage{xcolor}
\usepackage{amsmath}
\usepackage{amssymb}
\usepackage{xspace}
\usepackage[ruled,vlined,linesnumbered]{algorithm2e}

\SetKwComment{tcc}{//\ }{}
\SetArgSty{textnormal}
\SetAlgoSkip{}

\newcommand{\reals}{\mathbb{R}}
\newcommand{\rats}{\mathbb{Q}}
\newcommand{\ints}{\mathbb{Z}}
\newcommand{\nats}{\mathbb{N}}
\newcommand{\field}{\mathbb{F}}
\newcommand{\primefield}{\mathbb{Z}_p}
\newcommand{\st}{\;|\;}

\renewcommand{\span}[1]{\langle{#1}\rangle}

\newcommand{\sat}{\textsf{sat}\xspace}
\newcommand{\unsat}{\textsf{unsat}\xspace}
\newcommand{\unknown}{\textsf{unknown}\xspace}
\newcommand{\timeout}{{\textsf{t}/\textsf{o}}\xspace}

\newcommand{\civer}{\textsf{CIVER}\xspace}
\newcommand{\barcelogic}{\textsf{Barcelogic}\xspace}
\newcommand{\zeethree}{\textsf{Z3}\xspace}
\newcommand{\yices}{\textsf{Yices}\xspace}
\newcommand{\ffsol}{\textsf{ffsol}\xspace}
\newcommand{\maple}{\textsf{Maple}\xspace}
\newcommand{\cocoa}{\textsf{CoCoA}\xspace}
\newcommand{\cvc}{\textsf{cvc5}\xspace}

\usepackage{tcolorbox}
\usepackage{proof}

\begin{document}
\title{An Effective Orchestral Approach to Satisfiability Modulo Prime Fields}
%
%
\author{Miguel Isabel\inst{1}
  \and
 Enric Rodríguez-Carbonell\inst{2}
 \and
  Clara Rodríguez-Núñez\inst{1}
  \and
  Albert Rubio\inst{1}
}

%
%
\institute{Complutense University of Madrid, Spain \and
Technical University of Catalonia, Spain}
\maketitle              
\begin{abstract}
Zero-knowledge proofs (ZKPs) are an emerging technology that has become the solution to efficiently provide security and privacy along with the transparency requirement of blockchains. ZKPs are usually expressed by means of arithmetic circuits and, more generally, systems of polynomial equations in a large prime field (commonly ranging from 64-bit to 256-bit values).
  An increasing interest to apply formal verification techniques to ensure soundness and completeness properties of ZKP protocols has shown the need of developing powerful SMT solvers able to handle such constraint systems.
  In this paper we  consider the problem of deciding the satisfiability of existentially quantified first-order formulas defined over polynomial equations on a prime field. We present a new DPLL($T$)-based approach in which the theory solver orchestrates several modules with different trade-offs between completeness and efficiency. We have implemented the proposed techniques in a prototype that already shows better results than existing state-of-the-art tools on both benchmarks from the domain of ZKP compiler correctness and new benchmarks coming from the verification of arithmetic circuits for ZKPs.
\keywords{SMT \and Finite field \and Polynomials \and Zero-Knowledge Proofs.}
\end{abstract}

\section{Introduction}
\label{sec:introduction}

 A zero-knowledge proof (ZKP) is an emerging type of cryptographic protocol  that efficiently provides security and privacy while maintaining the transparency required in distributed ledgers like blockchains. A ZKP allows one party (the \emph{prover}) to prove to another party (the \emph{verifier}) that some statement is true, without revealing any information apart from the fact that it is true. In addition to being \emph{zero-knowledge} (it cannot leak any information besides the truth of the statement), a ZKP has to satisfy \emph{soundness}: if the statement is false, no dishonest prover can convince the honest verifier, except with some small probability; and \emph{completeness}: if the statement is true, an honest verifier will be convinced by an honest prover.
  ZKPs are usually expressed by means of arithmetic circuits and, more generally, systems of polynomial equations in a large prime field. Although small primes are used in some protocols, for security and efficiency reasons the size of the field commonly ranges from 64 bits to 256 bits.
  The intricacies of designing ZKP protocols greatly increases the probability of introducing vulnerabilities, whose presence in critical applications (proof of identity, interoperability, scalability, etc.) may compromise, for instance, funds or private data. For this reason, there is a growing interest in applying formal verification techniques to ensure soundness and completeness properties of ZKP protocols. To this end, the development of powerful SMT solvers able to handle polynomial equality constraint systems on a large prime field is crucial.
  Here we  consider the satisfiability problem of existentially quantified first-order formulas where atoms are polynomial equations on a prime field. With the aim of providing both efficiency and scalability, our main contribution is a new DPLL($T$)-based approach in which the theory solver orchestrates several modules with different tradeoffs between completeness and performance. In particular we have:

  \begin{itemize}

  \item \textbf{A Gröbner bases module.} This module uses Gröbner bases \cite{Cox+Others/1991/Ideals} to check if there is an inconsistency in the current assignment. Namely, it exploits the well-known result from algebraic geometry that ensures that if $1$ belongs to the ideal generated by the polynomials in a system of equations, then there is no solution. As the involved computations of Gr\"obner bases may be costly, in practice this module is only called at the leaves of the search tree.

\medskip

  \item \textbf{A prime field linear module.} On the opposite side of the spectrum, this module efficiently finds a solution to a system of linear constraints, or explains why such a solution cannot exist. By abstracting monomials with fresh variables, this module can also be applied to polynomial constraints, at the cost of sacrificing precision: while conflicts in the linear abstraction always carry over to the original constraints, a solution is valid only when it is compatible with the semantics of non-linear monomials.

\medskip    
    
\item \textbf{An equivalence inference module.} The goal of this module is to efficiently identify by means of union-find data structures whether in the current assignment it holds that $x_{1} = x'_{1} \;\land\; \cdots \;\land\; x_{n} = x'_{n} \;\land\; y = f(x_{1}, \ldots, x_{n}) \;\land\; y' = f(x'_{1}, \ldots, x'_{n})$ for some variables $x_{1}, \ldots, x_{n}$ and some polynomial $f$. If so, $y = y'$ can be derived. This frequently arises, e.g., when verifying that a given arithmetic circuit is safe \cite{Circom2023,DBLP:conf/sp/IsabelRR24,pldi23-picus}. 

\medskip
    
\item \textbf{An integer linear module.} This module analyzes the current assignment, detects those variables that have a domain smaller than the prime field $\primefield$ and collects those linear equations $E = 0$ and disequations $E \not= 0$ such that $E$ always takes values in $\{-p+1,\ldots,p-1\}$. Since no wrapping is possible, for that subset of constraints, unsatisfiability in $\ints$ implies unsatisfiability in $\primefield$, in which case an explanation is transferred to the SAT engine. Linear constraints that take values beyond this range can also be handled with fresh variables. Similar ideas were introduced in \cite{DBLP:conf/sp/IsabelRR24} in a once-and-for-all transformation, but now they are dynamically exploited in a theory solver.

\medskip
    
\item \textbf{A linear clause inference module.} Even when constraints are not linear, implied (and even equivalent) linear constraints can be inferred. E.g., the equation $x(x-1) = 0$ is commonly used to encode that a variable $x$ is binary, i.e., that $x = 0 \,\lor\, x = 1$. In general this module derives \emph{linear clauses}: clauses of linear constraints that are logical consequences of the formula. Similarly to how a lemma is learnt after a theory conflict, these redundant clauses are added to the SAT solver clause database.
  
\medskip
    
\item \textbf{A real non-linear module.} The unique feature of this module is that it is aimed at generating models. Since the only other component able to produce solutions is the one for linear constraints, we have added this module which works on $\reals$ instead of $\primefield$. Thus we can capture many instances with models over $\rats$ which, once translated, turn out to be models over $\primefield$ as well. Note that, in a field, all elements but zero have an inverse, so a rational number has an equivalent in $\primefield$ provided the denominator is not a multiple of $p$.

  \end{itemize}

Given the complexity of the theory under consideration, the price to be paid for our approach geared towards efficiency is the loss of completeness of the method: for some formulas it may not be able to prove or refute satisfiability, and hence the conclusion may be an \unknown verdict.
  
  In spite of these theoretical shortcomings, the method has been the implemented in a prototype that already shows better results than state-of-the-art tools both on existing benchmarks~\cite{DBLP:conf/cav/OzdemirKTB23} and on new ones coming from the verification of arithmetic circuits for ZKPs~\cite{DBLP:conf/sp/IsabelRR24}.

  The paper is structured as follows. In Section \ref{sec:preliminaries} we review basic concepts from algebra and SMT. Section \ref{sec:theory-solver} describes the modules of the theory solver and how they are coordinated. In Section \ref{sec:experiments} an evaluation of our prototype is shown, which validates the effectiveness of our techniques. Section \ref{sec:related-work} examines related work, and Section \ref{sec:conclusions-and-future-work} presents the conclusions and draws lines for future research.
  



\section{Preliminaries}
\label{sec:preliminaries}

\subsection{Algebra}
\label{sec:preliminaries:algebra}

A \emph{field} is a set equipped with two commutative operations,
called \emph{addition} and \emph{multiplication}, such that: (i) it is
a group under addition, where an element denoted by $0$ is the
additive identity; (ii) the elements different from $0$ form a group
under multiplication, where an element denoted by $1$ is the
multiplicative identity; and (iii) multiplication distributes over
addition. A \emph{finite field} is a field that is a finite set, and
its \emph{order} is its size. All finite fields have
order $q$ of the form $q = p^{m}$ for some prime number $p$ and
$m \in \nats$. Up to isomorphism, the field of order $q$ is
unique and denoted by $\field_{q}$. In particular, a field of
order $p$ is called a \emph{prime field} and is isomorphic to
$\ints_{p}$, the integers modulo $p$.

Given a finite field $\field$ and variables $x_{1}, \ldots, x_{n}$,
the set of polynomials in $x_{1}, \ldots, x_{n}$ with coefficients in
$\field$ is represented by $\field[x_{1}, \ldots, x_{n}]$. An
\emph{ideal} $I \subseteq \field[x_{1}, \ldots, x_{n}]$ is a subset of
polynomials that forms a subgroup of the additive group of
$\field[x_{1}, \ldots, x_{n}]$ and is closed under multiplication by
arbitrary polynomials.
Given a set of polynomials $S = \{f_{1} , \ldots, f_{k} \}$, the set $\span{S} = \{g_{1} f_{1} +
\ldots + g_{k} f_{k} \st g_{i} \in \field[x_{1} , \ldots, x_{n} ] \}$
is an ideal, called the \emph{ideal generated} by $S$. Then $S$ is
said to be a \emph{basis} of $\span{S}$. The \emph{ideal membership
problem} consists in, given a polynomial $f$, determining whether $f
\in \span{S}$.
This is related to the existence of solutions to the system of
equations $\{f_{1} = 0, \ldots, f_{k} = 0\}$: if $1 \in \span{S}$ then
the system has no solutions. In general the converse may not be true,
though.

Ideal membership can be algorithmically solved by means of a special kind of bases called \emph{Gröbner bases}.
For details we refer the reader to \cite{Cox+Others/1991/Ideals}.
Here it is enough to know that, if $S$ is a set of polynomials and $G$ is a Gröbner basis for the ideal $\span{S}$, then there is an algorithm (called \emph{reduction}) that determines if a given polynomial $f$ belongs to $\span{S} = \span{G}$.
Well-known algorithms for computing
Gr\"obner bases are Buchberger's
\cite{Cox+Others/1991/Ideals} and more modern Faugère's F4
\cite{FAUGERE_F4} and F5 \cite{FAUGERE_F5} algorithms. Unfortunately,
the worst-case complexity of computing a Gröbner basis for a
set of polynomials in $n$ variables and maximal degree $d$ is lower
bounded by $d^{2^{\Omega(n)}}$, i.e., at least doubly
exponential in $n$.

\subsection{Satisfiability Modulo Theories}
\label{sec:preliminaries:satisfiability-modulo-theories}

Given a theory $T$, a \emph{theory solver} for $T$ or \emph{$T$-solver} is a procedure for deciding if
a conjunction of literals of $T$ is satisfiable or not. Thanks to the
DPLL($T$) framework,
the satisfiability of a
formula built from atoms of $T$ and propositional
connectives can be decided by interweaving a propositional satisfiability
solver and a $T$-solver. In DPLL($T$), the $T$-solver maintains a
state that is an internal representation of the literals asserted so
far. This solver must provide operations for updating the state when
new literals are asserted, checking whether the state is consistent
with the theory, and backtracking. Optionally, the solver may also
implement theory propagation, that is, it may identify unasserted
literals that are implied in the theory by the current state. To
cooperate with the SAT engine, the theory solver must also
produce an explanation for each conflict, that is, an inconsistent
subset of the asserted literals, and an explanation for each
theory-propagated literal $\ell$, i.e. a subset $E$ of the asserted literals such that $E \models \ell$.


\section{Theory Solver}
\label{sec:theory-solver}

In this paper we address the problem of, given a formula built by applying propositional connectives to atomic polynomial equations over a prime field $\primefield$, determining if there is an assignment of values to variables such that the formula is satisfied. To that end a DPLL($T$) architecture will be adopted, which only requires a theory solver that is able to determine the satisfiability of a conjunction of theory literals, in this case polynomial equations and disequations (that is, negations of equations).
The goal of this section is to describe the design of the theory solver. First, each of the modules it consists of will be examined. Then it will be explained how these components should be orchestrated.

\subsection{Gr\"obner Bases Module}
\label{sec:groebner-bases-module}
To simplify the presentation, let us first assume that we need to check the consistency of a set of polynomial equations (i.e., there are no disequations). As reviewed in Section \ref{sec:preliminaries}, it is well-known that Gr\"obner bases are a handy tool for that \cite{DBLP:conf/cav/OzdemirKTB23,DBLP:conf/cav/OzdemirPBFBD24}. Still, implementing efficiently the algorithms for generating Gr\"obner bases is a highly non-trivial task. For this reason, our Gr\"obner bases module relies on an external program which is expected to offer the following features:

\begin{enumerate}

\item
\label{item:is-one-in-basis}
  given a set of polynomials $S = \{f_{1} , \ldots, f_{k} \}
  $, it determines if $1 \in
\span{S}$; and

\item
\label{item:certify-one-is-in-basis}
if $1 \in \span{S}$, it produces a \emph{certificate} of ideal
membership, that is, polynomials $g_{1} , \ldots, g_{k}
$ such that
$1 = g_{1} f_{1} + \ldots + g_{k} f_{k}.$
  
\end{enumerate}

The Gr\"obner bases module takes the conjunction of asserted
polynomial equations $f_{1} = 0 \land \ldots \land f_{k} = 0$ as
input. Feature 1 is used to detect when this conjunction is
unsatisfiable, as $1 \in \span{\{f_{1} , \ldots, f_{k} \}}$ implies
that there are no solutions. And if this condition holds, Feature 2
yields an explanation of inconsistency: one just needs to include all
those $f_{i} = 0$ such that the corresponding $g_{i}$ in the
certificate is not null. Interestingly, the certificate allows verifying the
results given by the external system, in case it is not trusted. On
the other hand, it must be pointed out that this technique is
\emph{sound} (if a set of literals is reported to be unsatisfiable, it
is indeed unsatisfiable), but \emph{incomplete}: field polynomials \cite{Hader2002MasterThesis} are
\emph{not} added to the asserted polynomials due to their degree, and therefore it may be that $1
\not\in \span{S}$ but there are no solutions. As a
consequence, if $1 \not\in \span{S}$ then
satisfiability cannot be guaranteed and the module returns
\unknown as a result. However, this shortcoming can be alleviated
with the interaction with the other modules, as will be explained in
Sections \ref{sec:prime-field-linear-module} and
\ref{sec:real-non-linear-module}.

Implementation-wise, Features \ref{item:is-one-in-basis} and
\ref{item:certify-one-is-in-basis} are supported off-the-shelf by
standard computer algebra tools. Feature \ref{item:is-one-in-basis}
can be achieved by computing a \emph{reduced Gr\"obner basis} \cite{Cox+Others/1991/Ideals} $G$ of $S$ and testing
if $G = \{1 \}$, as $1 \in \span{S}=\span{G} $ if and only if $G = \{1
\}$. And Feature \ref{item:certify-one-is-in-basis} is a particular
case of expressing the polynomials of $G$ as a
combination of the original generators $S$. Both computing a reduced Gr\"obner
basis and expressing its polynomials in terms of the original ones are
common functionalities that are available in many symbolic algebra
systems
\cite{maple,monagan2012maple,DBLP:conf/issac/BigattiR06,M2,Mathematica}.
Hence, our Gr\"obner bases module acts essentially as a wrapper for a
computer algebra tool that is taken as a black box without further
intervention. As will be seen in Section \ref{sec:experiments}, this
hands-off approach allows one to easily accommodate to different
computer algebra tools in a very flexible way.

Last but not least, in general some of the literals to be checked for consistency
may be polynomial \emph{disequations}. To handle those, the well-known
``Rabinowitsch trick'' \cite{Rabinowitsch1930} can be employed: a polynomial
disequation $f \neq 0$ is replaced by the polynomial equation $u f - 1
= 0$, where $u$ is a fresh variable. Note that for any $\alpha$ such that
$f(\alpha) \neq 0$ we may assign to $u$ the value $\frac{1}{f(\alpha)}$, which is well
defined and satisfies $u f - 1 = 0$; and vice versa, if $u(\alpha)\ f(\alpha) - 1 =
0$ then it must be $f(\alpha) \neq 0$, since if it were $f(\alpha) = 0$ then we would
get the contradiction $-1 = 0$.

\begin{example}
Let us assume that the asserted literals are $x \neq 0, x \neq 1, x (x-1) = 0$, and that the module is asked to determine their consistency\footnote{In the examples for which the precise prime field is not relevant, it will be omitted.}. Fresh variables $u$, $v$ are introduced so as to express the disequations $x \neq 0$ and $x \neq 1$ as equations, in this case $u x - 1 = 0$ and $v (x - 1) - 1 = 0$ respectively. A reduced Gr\"obner basis for $\{u x - 1, v (x - 1) - 1, x (x-1)\}$ is $\{ 1 \}$, which implies that there is a conflict: indeed, $x (x-1) = 0$ is equivalent to $x = 0 \,\lor\, x = 1$. Now, if in the ensuing search literal $x (x-1) = 0$ is retracted and the module is asked to check the consistency of $x \neq 0, x \neq 1$, a reduced Gr\"obner basis is computed for $\{u x - 1, v (x - 1) - 1\}$. Such a basis is $G = \{u v + u - v, v x - v - 1, u x - 1\}$ (with respect to the lexicographical order $x > u > v$). As $G \neq \{ 1 \}$, in this case the module would return \unknown.
\end{example}


\subsection{Prime Field Linear Module}
\label{sec:prime-field-linear-module}
In many instances of practical interest, most if not all of the constraints are linear.
Hence, it is likely that conflicts only involve linear constraints. And even if some non-linear constraint is part of a conflict, sometimes the semantics of monomials is not relevant. The following example illustrates this situation.

\begin{example}
  Let $S = \{ x - y^{2} \neq 0,\; y^{2} - y z = 0,\; y z - x = 0 \}$. This system is inconsistent: by adding the last two equations it can be deduced that $y^{2} - x = 0$, which contradicts the first constraint. Moreover, $S$ is not linear. Now, if monomials $y^{2}$ and $yz$ are abstracted by fresh variables $u$ and $w$ respectively, the resulting \emph{linear} system of constraints $S' = \{ x - u \neq 0,\; u - w = 0,\; w - x = 0 \}$ is still inconsistent, with a similar argument. Note that to determine the satisfiability of this linearized system, disequations (such as $x - u \neq 0$) do not need to be transformed into quadratic equations with the ``Rabinowitsch trick'', and instead can be handled using only simpler and more efficient linear reasoning.
\end{example}

Motivated by these observations, the prime field linear module was incorporated to the theory solver.
It takes as input the asserted constraints and may return as a verdict \sat (and a certifying model), \unsat (and a conflict explanation), or \unknown. Similarly to \cite{DBLP:conf/cav/DutertreM06} for real linear arithmetic, the module requires that the input formula is preprocessed before the search begins. Namely, for each atomic equation of the form $f(x) = k$, where $f(x)$ is a monic
polynomial without a constant term and $k \in \primefield$, in the first place the non-linear monomials in $f$ are \emph{abstracted}; that is, for any non-linear monomial $m$, a fresh auxiliary \emph{monomial variable} $v_{m}$ is introduced once and for all, and then all occurrences of $m$ in $f$ are replaced by $v_{m}$. As a result, $f$ is transformed into a linear polynomial, its \emph{abstraction}, which will henceforth be referred to as $g$. Now, if $g$ is not a variable, yet another variable is introduced: a fresh auxiliary \emph{slack variable} $s_{g}$, which is used to replace $g(x) = k$ by $s_{g} = k \;\land\; s_{g} = g(x)$. Altogether, the preprocess decomposes the input formula as $F \land G \land H$, where: $F$ is a CNF whose literals are of the form $x = k$ or $x \neq k$ (called \emph{domain constraints}), being $x$ a variable and $k \in \primefield$; $G$ is a conjunction of linear equations of the form $s_{g} = g(x)$; and $H$ is a conjunction of polynomial equations of the form $v_{m} = m$. The module essentially ignores $H$ (but not completely, see below), and takes $G$ as fixed during the search, so that only domain constraints from $F$ are dynamically asserted/retracted.

\begin{example}
  Let $F$ be $x - y^{2} \neq 0 \;\land\; y^{2} - y z = 0 \;\land\; (y z - x = 0 \,\lor\, x + y = 0 \,\lor\, z^{2} = 1).$ There are only three non-linear monomials in the formula $F$: $y^{2}$, $y z$ and $z^{2}$, which are abstracted by monomial variables $v_{y^{2}}$, $v_{y z}$ and $v_{z^{2}}$, respectively. So the formula can be equivalently expressed as
\begin{multline*}
  x - v_{y^{2}} \neq 0 \;\;\land\;\; v_{y^{2}} - v_{y z} = 0 \;\;\land\;\; (v_{y z} - x = 0 \;\lor\; x + y = 0 \;\lor\; v_{z^{2}} = 1)\\ \;\;\land\;\; v_{y^{2}} = y^{2} \;\;\land\;\; v_{y z} = y z \;\;\land\;\; v_{z^{2}} = z^{2}.
\end{multline*}  
%
Linear polynomials $x - v_{y^{2}}$, $v_{y^{2}} - v_{y z}$, $v_{y z} - x$ and $x+y$ can be replaced by slack variables $s_{x - v_{y^{2}}}$, $s_{v_{y^{2}} - v_{y z}}$, $s_{v_{y z} - x}$ and $s_{x+y}$, respectively. This allows us to finally represent the formula as $F \land G \land H$, where
  $F$ and $G$ are

\smallskip
  
  $
  \hspace*{-0.5cm} \begin{array}{l}
      s_{x - v_{y^{2}}} \neq 0 \,\,\land\,\, s_{v_{y^{2}} - v_{y z}} = 0 \,\,\land\,\, (s_{v_{y z} - x} = 0 \,\lor\, s_{x+y} = 0 \,\lor\, v_{z^{2}} = 1)\hbox{ and } \\
      s_{x - v_{y^{2}}} = x - v_{y^{2}} \,\,\land\,\, s_{v_{y^{2}} - v_{y z}} = v_{y^{2}} - v_{y z} \,\,\land\,\, s_{v_{y z} - x} = v_{y z} - x \,\,\land\,\, s_{x+y} = x+y
    \end{array}
    $

\smallskip

  \noindent
  respectively, and $H$ is $v_{y^{2}} = y^{2} \;\land\; v_{y z} = y z \;\land\; v_{z^{2}} = z^{2}$.
  
\end{example}

In what follows, by variables we will mean both original and auxiliary (monomial or slack) variables. The main data structures that define the state are:

\begin{itemize}

\item A stack with the asserted literals, called the \emph{trail}.

\item A map $\sigma$ that assigns a value $k \in \primefield$ to each variable $x$, called the \emph{solution}.

\item A map that assigns to each variable $x$ the index in the trail of a literal of the form $x = k$ where $k \in \primefield$, if there is any (in this case we say that $x$ is \emph{fixed}).

\item A map that assigns to each variable $x$ the stack of indices in
the trail of the literals of the form $x \neq k$, for a certain $k \in
\primefield$.

\item A \emph{tableau}, which is a system of equations in solved form that is equivalent to the $G$ part of the (preprocessed) input formula $F \land G \land H$. Solved variables will be called
\emph{basic} variables, while the others will be \emph{non-basic}.
This terminology is inherited from the simplex algorithm
\cite{Schrijver1998}. Note that, in the solution $\sigma$, the values of the non-basic variables determine those of the basic variables.

\end{itemize}

As an invariant during the search, it is preserved that the solution $\sigma$ satisfies both the tableau and the domain constraints of non-basic variables.
When the module is required to check the consistency of the asserted literals, as in \cite{DBLP:conf/cav/DutertreM06}, first a lightweight check is performed with each of the literals, in chronological order. If the literal is an equation of the form $x = k$, this lightweight check first tests whether there is a trivial conflict (with another equation of the form $x = l$ where $k \neq l$, or with the disequation $x \neq k $); if that is not the case, and if $x$ is non-basic, the solution $\sigma$ is updated so that $\sigma(x) = k$. Otherwise, if the literal is a disequation of the form $x \neq k$, the lightweight check first tests whether there is a trivial conflict (with the equation $x = k$); if that is not the case, and if $x$ is non-basic and its current value in the solution $\sigma$ is $k$, to maintain the invariant a new value for $\sigma(x)$ is sought which is compatible with the domain constraints of $x$, and $\sigma$ is updated accordingly. Once the lightweight checks are over and if no conflict has been detected yet, the heavier consistency check in Algorithm \ref{alg:solver} (which is an adaption of the \textsf{Check} procedure from \cite{DBLP:conf/cav/DutertreM06}) is invoked.

\begin{algorithm}[h]
\DontPrintSemicolon
\SetAlgoNoEnd
\SetAlgoLined
\caption{Heavy Consistency Check}
\label{alg:solver}

$\textit{result} \gets \unknown$\;

\While{$\textit{result} = \unknown$}{
    $x \gets$ Find a basic variable such that $\sigma$ violates its domain constraints\;
    
    \If{$x \neq \bot$}{
        \uIf{$x$ has a domain constraint $x = k$ violated by $\sigma(x)$}{
            $y \gets$ Find a non-fixed non-basic variable in the equation of $x$\;
            \If{$y \neq \bot$}{
                Swap $x$ and $y$ in the tableau, and update $\sigma$ so that $\sigma(x) = k$\;
            }\lElse{                
                $\textit{result} \gets \unsat$
            }
          }\Else(\tcp*[h]{$x$ has a domain constraint $x \neq k$ violated by $\sigma(x)$}){
            $y \gets$ Find a non-fixed non-basic variable in the equation of $x$\;
            \If{$y \neq \bot$}{
              Find value $l$ for $y$ compatible with the disequalities of $y$ that determines a value for $x$ compatible with the disequalities of $x$, and update $\sigma$ so that $\sigma(y) = l$\;
            }\lElse{
                $\textit{result} \gets \unsat$
            }
        }
    }\lElse{
        $\textit{result} \gets \sat$
    }
}
\Return $\textit{result}$\;
\end{algorithm}

If the outcome of this call  is that a
conflict was found (that is, \unsat is returned), an explanation is
compiled. Namely, if the procedure returns at line 9, then the
explanation for the conflict consists of all literals that fix the
non-basic variables in the equation of $x$, and also the violated domain constraint $x = k$. And similarly, if the procedure
returns at line 14, then the explanation consists of
all literals that fix the
non-basic variables in the equation of $x$, and also the violated domain constraint $x \neq k$.
Otherwise, i.e., if \sat is returned, the solution $\sigma$ is a model
of the linearized formula. More precisely, if $M$ is the conjunction
of the linear abstractions of the asserted literals, then $\sigma$
satisfies $M \land G$, where $G$ is the system of linear equations of
the input formula $F \land G \land H$. However, $\sigma$ may
not respect the semantics of non-linear monomials, i.e., it may not
satisfy $H$. For this reason, only when the search reaches a leaf,
that is, $M \models F$, it is eventually checked that $\sigma$
satisfies $H$. If that is the case, and therefore $\sigma \models F
\land G \land H$, the formula is reported to be satisfiable.
Otherwise, an \unknown verdict is issued.

Besides consistency checks, the module also performs limited theory propagation. Disequalities are propagated using that $x = k$ implies $x \neq l$ for all $l \neq k$. Moreover, if monomial variable $v_{m}$ corresponds to monomial $m$ and $x$ is a variable in $m$, then $x = 0$ implies $v_{m} = 0$. Another simple case occurs when all variables are fixed: if $m$ is the monomial $x_{1}^{e_{1}} \cdots x_{n}^{e_{n}}$, and $x_{1}, \ldots, x_{n}$ are fixed to values $k_{1}, \ldots, k_{n}$ respectively, then one can propagate $v_{m} = k_{1}^{e_{1}} \cdots k_{n}^{e_{n}}$. Note that these propagations exploit the non-linear semantics of monomials, and open the door for the rest of the module to take that into account. Finally, if in an equation implied by the system of linear equations $G$ all variables are fixed but one, then that one automatically gets a fixed value that can be propagated.


\subsection{Equivalence Inference Module}
\label{sec:equivalence-inference-module}
A key issue in the verification of ZKPs protocols is the \emph{safety problem} for arithmetic circuits \cite{Circom2023}: to check that, for a given circuit, the outputs are uniquely determined by the inputs. This yields instances of the following form: if $C(i, t, o)$ is a set of polynomial constraints defining a circuit with input signals $i = (i_{1}, \ldots, i_{m})$,
intermediate signals $t$ and output signals $o = (o_{1}, \ldots, o_{n})$, the formula
$$C(i, t, o) \;\land\; C(i', t', o') \;\land\; i_{1} = i'_{1} \;\land\; \cdots \;\land\; i_{m} = i'_{m} \;\land\; (o_{1} \neq o'_{1} \;\lor\; \cdots \;\lor\; o_{n} \neq o'_{n} )$$
is satisfiable if and only if the circuit is unsafe.

\begin{example}
  Let us consider a circuit with inputs $i_{1}$ and $i_{2}$, intermediate signals $t_{1}$ and $t_{2}$ and outputs $o_{1}$ and $o_{2}$, described by the set of polynomial equations $\{ t_{1} = i_{1}^{2},\; t_{2} = i_{2}^{2},\; o_{1} = t_{1} + t_{2},\; o_{2} = t_{1} - t_{2} \}\, .$
  That is, given inputs $i_{1}$ and $i_{2}$, the circuit computes as outputs $o_{1} = i_{1}^{2} + i_{2}^{2}$ and $o_{2} = i_{1}^{2} - i_{2}^{2}$.
  In this case, the verification of the safety of the circuit amounts to proving that the formula
  $$
  \begin{array}{lclclc}
\phantom{\land\;} t_{1}  = i_{1}^{2}    & \;\land\; & t_{2}  =  i_{2}^{2}    & \;\land\; & o_{1} = t_{1} + t_{2} \;\land\; o_{2} = t_{1} - t_{2} \\
    \land\; t'_{1} = {i'}_{1}^{2} & \;\land\; & t'_{2} =  {i'}_{2}^{2} & \;\land\; & o'_{1} = t'_{1} + t'_{2} \;\land\; o'_{2} = t'_{1} - t'_{2} \\
    \land\; i_{1} = i'_{1}        & \;\land\; & i_{2} = i'_{2}        & \land\; & (o_{1}  \neq  {o'}_{1} \;\lor\; o_{2}  \neq  {o'}_{2})
  \end{array}
  $$
  is unsatisfiable. And indeed it is so, since:

  \begin{enumerate}
  \item $i_{1} = i'_{1}$, $t_{1} = i_{1}^{2}$ and $t'_{1} = {i'}_{1}^{2}$ imply $t_{1} = t'_{1}$;
  \item $i_{2} = i'_{2}$, $t_{2} = i_{2}^{2}$ and $t'_{2} = {i'}_{2}^{2}$ imply $t_{2} = t'_{2}$;
  \item $t_{1} = t'_{1}$, $t_{2} = t'_{2}$, $o_{1} = t_{1} + t_{2}$ and $o'_{1} = t'_{1} + t'_{2}$ imply $o_{1} = o'_{1}$; and
  \item $t_{1} = t'_{1}$, $t_{2} = t'_{2}$, $o_{2} = t_{1} - t_{2}$ and $o'_{2} = t'_{1} - t'_{2}$ imply $o_{2} = o'_{2}$.
    
  \end{enumerate}
  
\end{example}

As shown in the example, efficient propagation of equivalences is crucial when handling this kind of problems. Inspired by existing work on the theory of EUF \cite{DBLP:conf/lpar/NieuwenhuisO03} we have incorporated this module, which by means of union-find data structures \cite{Cormen} tries to apply the next rule:

\medskip

\hspace{0.3cm}
\infer
    {%
      y = y'
    }
    {x_{1} = x'_{1} \;\land\; \cdots \;\land\; x_{n} = x'_{n} \;\land\; y = f(x_{1}, \ldots, x_{n}) \;\land\; y' = f(x'_{1}, \ldots, x'_{n})}

\medskip
    
\noindent
where $x_{1}, \ldots, x_{n}$ are variables and $f$ stands for an arbitrary polynomial.


\subsection{Integer Linear Module}
\label{sec:integer-linear-module}
This module is introduced to mitigate the difficulties of reasoning with inequations in a prime field. Although, strictly speaking, in our theory there is no order relation $\leq$ but only equality, univariate polynomials are sometimes used to express membership to a range. For instance, $x(x-1)=0$ is equivalent to $0 \leq x \leq 1$. In the presence of constraints like this, neither the Gröbner bases module nor the
prime field linear module can take advantage of the implied bounds. However, a linear integer arithmetic (LIA) solver can efficiently exploit such information to detect conflicts, as the following example illustrates.

\begin{example}
  Let us consider the following assignment of literals in $\ints_{p}$:\\
        \begin{minipage}{0.27\linewidth}
          $$
          \begin{array}{lcl}
            x_0  (x_0 - 1) = 0\\
            \dots\\
            x_{n-1}  (x_{n-1} - 1) = 0 \\ 
          \end{array}
          $$
        \end{minipage}
        \begin{minipage}{0.27\linewidth}
          $$
          \begin{array}{lcl}
            y_0  (y_0 - 1) = 0\\
            \dots\\
            y_{n-1}  (y_{n-1} - 1) = 0
          \end{array}
          $$
        \end{minipage}
        \begin{minipage}{0.46\linewidth}
          $$
          \begin{array}{lcl}
            \mathit{in} = x_0 + 2 x_1 + \dots + 2^{n-1}  x_{n-1} \\
            \mathit{in} = y_0 + 2 y_1 + \dots + 2^{n-1}  y_{n-1}
          \end{array}
          $$
        \end{minipage}
        \medskip
        
        \noindent
        Now, if we add any negated literal $ x_i \neq y_i$ with $i\in\{0,\ldots,n-1\}$, the resulting assignment is inconsistent in $\ints_{p}$ with $p>2^n-1$, since either $\mathit{in}$ cannot be expressed in binary with $n$ bits (when $\mathit{in} \geq 2^n$) or there is a single way to express it (so no bit $x_i$ and $y_i$ can be different when expressing the same number $\mathit{in}$). However, Gröbner bases struggle to prove this even for a very small $n$. Similarly, the modules from Sections \ref{sec:prime-field-linear-module} and \ref{sec:equivalence-inference-module} for linear constraints do not help much, as one ends up enumerating all possible assignments of $x_j=0$ or $x_j=1$ (and the same for the $y$'s), which again is only feasible for small values of $n$. Instead, assuming $p>2^n-1$, we can formulate an equivalent problem in LIA:\\
        \begin{minipage}{0.25\linewidth}
          $$
          \begin{array}{lcl}
            0\leq x_0 \leq 1\\
            \dots\\
            0\leq x_{n-1} \leq 1 \\ 
          \end{array}
          $$
        \end{minipage}
        \begin{minipage}{0.25\linewidth}
          $$
          \begin{array}{lcl}
            0\leq y_0 \leq 1\\
            \dots\\
            0\leq y_{n-1} \leq 1 \\ 
          \end{array}
          $$
        \end{minipage}
        \begin{minipage}{0.5\linewidth}
          $$
          \begin{array}{lcl}
            \mathit{in} - x_0 - 2 x_1 - \dots - 2^{n-1}  x_{n-1} = 0 \\
            \mathit{in} - y_0 - 2 y_1 - \dots - 2^{n-1}  y_{n-1} = 0 \\
            x_i < y_i  \;\;\lor\;\; x_i > y_i
          \end{array}
          $$
        \end{minipage}
        \medskip

        \noindent which is doable even for values of $n$ such that $2^n < p < 2^{n+1}$. Note that, in this case, one does not need to add the modulo operations, since the bounds on $x_0,\ldots, x_{n-1}, y_0,\ldots, y_{n-1}\in\{0,1\}$ and $\mathit{in}\in\{0,\ldots,p-1\}$
        imply that both  $\mathit{in} - x_0 - 2 x_1 - \dots - 2^{n-1}  x_{n-1}$ and $\mathit{in} - y_0 - 2 y_1 - \dots - 2^{n-1}  y_{n-1}$ are in $\{-p+1,\ldots,p-1\}$ when $p>2^n-1$, so no wrapping is possible. Thus any model in $\ints_{p}$ is a model in $\ints$, and therefore proving unsatisfiability in $\ints$ implies unsatisfiability in $\ints_{p}$.

\end{example}

Following the intuition given in the example, in general we try to detect lower and upper bounds on the variables. Then, as in \cite{DBLP:conf/sp/IsabelRR24}, these bounds are employed to detect linear constraints $E = 0$ or $E \not= 0$ in the current assignment whose expression $E$ takes values in $\{-p+1,\ldots,p-1\}$. These equations (or disequations) can then be passed to the LIA problem without any further transformation: the only case in which $E$ evaluates to $0$ in $\ints_p$ is when it is exactly $0$ (as it only takes values in $\{-p+1,\ldots,p-1\}$), hence any model in $\ints_p$ is also a model in $\ints$.  In case we cannot ensure that $E$ is in $\{-p+1,\ldots,p-1\}$, we can only use the equation $E = 0$ if we introduce a fresh integer variable $z$ and replace it by $E = z \cdot p$ instead, which is the equivalent equation in $\ints$. In the presence of disequations $E \not= 0$, with $E$ not bounded in $\{-p+1,\ldots,p-1\}$, we have to introduce two fresh integer variables $z_0$ and $z_1$ and replace the disequation by the equation $E = z_1 \cdot p + z_0$ together with $0 < z_0 < p$.

\subsubsection{Non-overflowing Linear Constraints}
\label{sec:non-overflowing-linear-constraints}

As explained above, the integer linear module attempts to detect lower and upper bounds on variables. Let us introduce now the deduction rules applied by the module to derive bounds. We denote by $L(E)$ and $U(E)$ the lower and upper bounds of an expression $E$, respectively.



\begin{center}
	\begin{minipage}{0.45\textwidth}
		\centering

    \hspace{0.5cm}\infer
    {\alpha_1\leq L(x) \;\;\land\;\; U(x)\leq \alpha_n}
    {%
      (x - \alpha_1)  \cdots  (x - \alpha_n) = 0 & \alpha_1 \leq \dots \leq \alpha_n
    }

	\end{minipage}
\hfill
	\begin{minipage}{0.45\textwidth}

    \hspace{0.5cm}\infer
    {L(E) \leq L(x)  \;\;\land\;\; U(x)\leq U(E)}
    {%
      x - E = 0
    }
	\end{minipage}
  \end{center}


Initially, we consider that the only bounds of the variables are the trivial ones: $0\leq x \leq p-1$. The deduction rules are then applied iteratively until a fixed point is reached and no additional bounds can be inferred.

Once the bounds are calculated, the module identifies which linear constraints $E = 0$ or $E \neq 0$ have $E$ bounded in $\{-p+1,\ldots,p-1\}$ by checking if $- p + 1 \leq L(E) \land U(E)\leq p-1$. The constraints that pass the test, which we call \emph{non-overflowing}, are inserted into a set, the \emph{non-overflowing set}.

\subsubsection{Building the Conflict Explanation}

When the LIA solver returns \unsat, the theory solver has to provide the SAT solver with an explanation. This requires care, since the explanation should be expressed in terms of the original literals of the input formula in $\primefield$. Therefore, in the process of finding bounds and detecting non-overflowing constraints, the following data have to be gathered:
\begin{enumerate}
\item For every lower and upper bound, the literals used to infer them.
\item For every non-overflowing constraint, the literals used to infer it is non-overflowing, and the literal of the constraint itself.
\end{enumerate}
This information allows explaining the literals that appear in the LIA formula in terms of the literals in $\ints_p$. Hence once we get an \unsat answer, we take an unsatisfiable core $\mathit{UC}$ as the explanation of the unsatisfiability in LIA, and transform it into a conflict explanation in $\ints_p$ by replacing every literal in $\mathit{UC}$ by its explanation in $\ints_p$.


\subsection{Linear Clause Inference Module}
\label{sec:linear-clause-inference-module}


Many real-world constraint systems contain non-linear constraints from which implied linear clauses can be inferred.
This is common in constraint systems used in ZKP protocols where only arithmetic constraints are allowed. For instance, ZKP protocols usually encode a disjunction $A = 0 \lor B = 0$ as the polynomial constraint $A \cdot B = 0$.
In this case, non-linear constraints are hiding linear structures, complicating reasoning about them.  As we show in the following example, the inference of the implied linear clauses simplifies the detection of conflicts.

\begin{example}
	Let us consider the formula $x y + x =0 \;\land\; x \neq 0 \;\land\;  y + 1 \neq 0\, .$
	The first constraint can be factored as $x(y + 1) = 0$, which is equivalent to $x = 0 \lor y + 1 = 0$. So the resulting formula is linear and can be solved by the prime field linear module, without requiring the use of Gr\"obner bases or other non-linear reasoning techniques.

\end{example}

In order to enhance the performance of the solver, a module for detecting new linear clauses was introduced. This module implements the following simple deduction rules, which infer linear implied clauses from non-linear constraints.
We denote variables by $x_i$, linear expressions by $E_i$ and constants by $\alpha_i$:


\begin{center}
	\begin{minipage}{0.3\textwidth}
		\centering
		
			\hspace{1cm}\infer
		{E_1 = 0\;\;\lor\;\; E_2 = 0}
		{%
			E_1 \cdot E_2 = 0
		}
		
	\end{minipage}
	\hfill
	\begin{minipage}{0.65\textwidth}
		
				\centering
		\hspace{1cm}\infer
		{x_1 - \alpha_1 = 0 \;\;\lor\;\; \dots \;\;\lor\;\; x_n - \alpha_n = 0}
		{%
			(x_1 - \alpha_1) \cdots (x_n - \alpha_n) = 0
		}
		
	\end{minipage}

\medskip

	\begin{minipage}{0.65\textwidth}
\hspace{1cm}  \infer
{%
	x = 0 \;\;\lor\;\; E_1 + \dots + E_n = 0
}
{x \cdot E_1 + \dots + x \cdot E_n = 0}
\end{minipage}
\end{center}

%
%
%
%


In order to apply the first rule (top left), the module includes an algorithm for factoring quadratic polynomials over prime fields. This algorithm relies on the Tonelli-Shanks method \cite{shanks1973five} for computing square roots in $\primefield$. The second rule (top right) is purely syntactic and can only be applied in constraints expressed as a product of roots.

%
%
%
%
%
%
%
%
%
%
%
%
%
%
%

\subsubsection{Extended Non-overflowing Linear Constraints}

In this section we show how some more non-overflowing constraints can
be inferred, so as to avoid introducing fresh variables when the test
of Section \ref{sec:non-overflowing-linear-constraints} fails. The
following example illustrates the situation we try to cover.

\begin{example}
	Let us consider the following constraint system in $\ints_p$:
	
        \begin{minipage}{0.27\linewidth}
          $$
          \begin{array}{lcl}
            x_0  (x_0 - 1) = 0\\
            \dots\\
            x_{n-1}  (x_{n-1} - 1) = 0 \\ 
          \end{array}
          $$
        \end{minipage}
        \begin{minipage}{0.27\linewidth}
          $$
          \begin{array}{lcl}
            y_0  (y_0 - 1) = 0\\
            \dots\\
            y_{n-1}  (y_{n-1} - 1) = 0
          \end{array}
          $$
        \end{minipage}
        \begin{minipage}{0.46\linewidth}
          $$
          \hspace*{-0.4cm}\begin{array}{lcl}
            \mathit{in}_1 + \mathit{in}_2 = x_1 + \dots + 2^{n-1}  x_n  \\
	    \mathit{in}_1 + \mathit{in}_2 = y_1 + \dots + 2^{n-1}  y_n\\
            x_i \neq y_i \mbox{ for some $i\in\{0\ldots n-1\}$}
          \end{array}
          $$
        \end{minipage}
        \medskip
        
  In this case, the constraints $\mathit{in}_1 + \mathit{in}_2 = x_1 + \dots + 2^{n-1}  x_n$ and $\mathit{in}_1 + \mathit{in}_2 = y_1 + \dots + 2^{n-1}  y_n$ are not non-overflowing individually even if  $p > 2^n - 1$, as the sum $\mathit{in}_1 + \mathit{in}_2$ may exceed $p$. However, their difference:
  $$0 = x_1 + \dots + 2^{n-1}  x_n- y_1 -\dots - 2^{n-1}  y_n$$
  is implied by the original formula, and is non-overflowing if $2^n - 1<p$. Moreover, this constraint together with the bounds suffices to prove unsatisfiability.

  \end{example}

	The following rule generalizes the reasoning described above. Here $W$ denotes any expression (linear or non-linear) and we do not require it to be bounded.

        \medskip
        
	\hspace{0.2cm}\infer
	{E_1 - E_2 = 0}
	{%
		W + E_1 = 0 & W + E_2 = 0& -p + 1\leq L(E_1-E_2) & U(E_1-E_2)\leq p-1
	}


\medskip

To apply the rule, the system checks if there are constraints that have the same unbounded part, and if so, it computes their difference. If the resulting constraint is non-overflowing then it is also added to the non-overflowing set. Note that even when $W$ is linear, using this rule is better in general than adding the two overflowing equations with their respective fresh variables. In fact, our experiments have shown that overflowing constraints do not help in general, and the overhead of using them slows down the solver. For this reason, our current implementation only uses non-overflowing linear constraints.

\subsection{Real Non-Linear Module}
\label{sec:real-non-linear-module}
As in Section \ref{sec:integer-linear-module}, here we take advantage of reasoning in a theory different from $\ints_p$. However, this module is aimed at generating models. The motivation for introducing it in the toolkit is that the only other component able to produce solutions is the prime field linear module, and hence our chances to find models for satisfiable problems are rather modest.
In this case, we will stay in the non-linear polynomial fragment, but will work over $\reals$ instead of $\ints_p$. The reason to do so is two-folded. In the first place, solving polynomials over the reals is, in principle, simpler than solving them over the integers. And secondly, since in a field all elements but zero have an inverse, any rational number has an equivalent in the prime field as long as the denominator is not a multiple of the prime. By exploiting this observation we can capture many instances with models over $\rats$ because, once translated, they turn out to be models over $\primefield$ as well. 

As alternatives we have  also considered to formulate the problem over $\ints$ instead of $\reals$, and also introducing fresh variables to encode the modulo operation in equalities and disequalities (as explained in Section~\ref{sec:integer-linear-module}). Nevertheless, in all cases the results were weaker than just using polynomial reasoning over $\reals$.




Let us see how the model for $\ints_p$ is built. In a nutshell,
if the solver over the reals returns \sat, then we transform the obtained model into a potential solution over $\ints_p$ if possible, and if so we check if it is indeed a model for $\ints_p$.
The transformation applies the following cases for every value in the model over $\reals$:
\begin{itemize}
\item The value is a rational number. Then if the denominator is $0$ in $\ints_p$ we discard the model, otherwise we multiply the numerator by the inverse of the denominator in $\ints_p$.
\item The value is of the form $\sqrt{\alpha}$ for a certain $\alpha \in \rats$. Then we check if the $\ints_p$ equivalent to $\alpha$ has a square root in $\ints_p$. Otherwise, we discard the model.
\end{itemize}

If the translation is feasible for all values, then we check if it satisfies all literals in the current assignment, and if so we return \sat and the model.
To increase the chances of success, we call the solver over $\reals$ on two different sets of constraints: (i) the current assignment; and (ii), the equations of the (reduced) Gröbner basis computed by the Gröbner bases module.







\subsection{Orchestration}
\label{sec:orchestration}
Finally let us describe how the components the theory solver consists of are orchestrated. Given their different tradeoffs between completeness and performance, the natural criterion is to prioritize the modules according to their efficiency. Thus, when the theory solver is asked to check the consistency of the asserted literals, first of all the equivalence inference module is called. If an inferred equality between variables contradicts an asserted or theory propagated literal, a conflict is flagged and an explanation is returned. Otherwise, if some equality has been inferred, the control is given back to the SAT engine, so that unit propagation can be applied.

Otherwise, i.e., if the equivalence inference module does not derive anything, the prime field linear module is next. Again, if a conflict is discovered then an explanation is returned and the flow goes back to the SAT solver. Otherwise, if the current solution of the module is compatible with the semantics of monomials and is hence a true model of the asserted literals, and the search has reached a leaf, then the input formula is reported to be satisfiable.

Otherwise, that is, if the prime field linear module does not find a conflict and either a leaf has not been reached or the current solution is spurious, the integer linear module is invoked. Just as with the previous module, if a conflict is detected, then an explanation is passed to the SAT engine.

Otherwise, i.e., when the integer linear module does not find a conflict, it is the turn of the linear clause inference module. Now, if an inferred linear clause is false under the current literal assignment, an explanation for the conflict is returned to the SAT solver. In any case, if some linear clause has been inferred, the flow goes back to the SAT engine.

Finally, and only if a leaf of the search tree has been reached, as a last resort the Gröbner bases module is called. Again, if a conflict is encountered then an explanation is reported to the SAT solver. Otherwise, the conjunction of asserted literals is highly likely to be satisfiable, though it cannot be guaranteed as there is no witness for certifying it. As by assumption the search has reached a leaf and no more progress can be made, before eventually quitting the real non-linear module is called. This module is costly but is precisely aimed at finding models, and at this stage no other alternative is available before giving an \unknown result.

To conclude this section, Figure \ref{fig:orchestration} shows a diagram that illustrates graphically the orchestration of the modules in the theory solver that has just been described.
\begin{figure}[t]
  \begin{center}
    \scalebox{0.3}{\includegraphics{diagram.eps}}
  \end{center}
  \caption{Diagram of how the modules in the theory solver are orchestrated.}
  \label{fig:orchestration}
\end{figure}


\section{Experiments}
\label{sec:experiments}


We have implemented the proposed techniques as a theory solver of the SMT system \barcelogic \cite{DBLP:conf/cav/BofillNORR08}. The resulting SMT solver for the theory of prime fields will be henceforth called \ffsol. As regards external tools, for the Gr\"obner bases module we have implemented a plug-in for \maple \cite{maple}, which is commercial software, as well as for \cocoa \cite{DBLP:conf/issac/BigattiR06}, which is free. Besides, for both the integer linear and the real non-linear modules we use \zeethree \cite{DBLP:conf/tacas/MouraB08} version 4.16, as it is one of the best SMT solvers for the involved theories of Linear Integer Arithmetic (LIA) and Non-linear Real Arithmetic (NRA).

In order to assess the value of our contribution, we compare the performance of our solver against \cvc \cite{DBLP:conf/tacas/BarbosaBBKLMMMN22} version 1.2.2, the state-of-the-art SMT solver for the theory of finite fields, and \yices \cite{DBLP:conf/ijcar/HaderKIGK24} version  2.7.0. With the aim of performing a fair comparison, in the experiments reported below we chose \cocoa for the Gr\"obner bases module, since this is the computer algebra system that \cvc employs too.
For the evaluation we
consider two sets of instances. The first one was proposed in \cite{DBLP:conf/cav/OzdemirKTB23} and comes from the domain of ZKP compiler correctness. Thus, we can straightforward compare our system against \cvc and \yices. 
For the second benchmark set we employed the tool in \cite{DBLP:conf/sp/IsabelRR24}, which produces
a different class of instances arising as subproblems in the verification of arithmetic circuits.
The experimental results were obtained using an AMD Ryzen Threadripper PRO 3995WX 64-Core Processor with 512GB of RAM (Linux Kernel Debian 5.10.70-1) and a timeout of 300 seconds per instance. The main results are presented in the following sections.

\subsection{Comparison with Benchmarks from \cite{DBLP:conf/cav/OzdemirKTB23}}
\label{exp:cvc5}
We have compared our solver against \cvc and \yices on 1602 instances from the benchmark set in \cite{DBLP:conf/cav/OzdemirKTB23}. Table \ref{fig:cvc5benchmarks} shows how the two solvers, \ffsol (columns) and \cvc (rows), agree or differ on instance outcomes (\sat, \timeout, \unsat). Each cell in the table indicates the number of cases that correspond to a specific combination of results. For instance, both \cvc and \ffsol return \sat on 761 of the benchmarks. However, \ffsol identifies 50 satisfiable problems for which \cvc times out, while \cvc identifies only 28 satisfiable problems for which \ffsol times out.  A comparison of the execution times for these benchmarks reveals that \ffsol proves satisfiability on average in 0.7 seconds, whereas \cvc requires 1.5 seconds. The mean execution time for the 50 problems solved exclusively by \ffsol is 0.55 seconds. Moreover,  \cvc only solves  83.4\% of the instances whereas \ffsol solves 92.4\% of them.  It is important to point out that the Gröbner bases module is only invoked in 2\% of the unsatisfiable instances and 0.4\% of the satisfiable ones.
Regarding the results in Table \ref{fig:cvc5benchmarks-yices}, \yices identifies 59 \sat instances that \ffsol is unable to solve, while \ffsol finds 68 instances that \yices does not. This suggests that the techniques from \cite{DBLP:conf/ijcar/HaderKIGK24} could significantly enhance our solver for the \sat cases. Nevertheless, \yices solves only 66.2\% of the total instances.
\begin{center}
	\begin{minipage}{0.4\textwidth}
		\centering
\begin{tabular}{c|ccc} \diagbox{{\cvc}}{{\ffsol}} & {\sat} & {\timeout} & {\unsat} \\ \hline {\sat} & 761 & 28 & - \\ {\timeout} & 50 & 80 & 131 \\  {\unsat} & - & 14 & 538 \\ \end{tabular}
	\captionof{table}{Comparative results between \ffsol (columns) and \cvc (rows)}
	\label{fig:cvc5benchmarks}
	\end{minipage}
\hfill
\begin{minipage}{0.4\textwidth}
	\centering
\begin{tabular}{c|ccc} \diagbox{{\yices}}{{\ffsol}} & {\sat} & {\timeout} & {\unsat} \\ \hline {\sat} & 743 & 59 & - \\ {\timeout} & 68 & 62 & 408 \\  {\unsat} & - & 1 & 261 \\ \end{tabular}
	\captionof{table}{Comparative results between \ffsol (columns) and \yices (rows)}
	\label{fig:cvc5benchmarks-yices}
	\end{minipage}
\end{center}

\subsection{Comparison with Benchmarks from \cite{DBLP:conf/sp/IsabelRR24} }
\label{exp:civer}
 In order to evaluate our solver’s performance on a different class of benchmarks, namely coming from the verification of arithmetic circuits, we have instrumented the \civer tool \cite{DBLP:conf/sp/IsabelRR24} to obtain new instances. \civer is a verification tool for proving a significant property in arithmetic circuits known as \emph{weak safety} (also known as determinism), which basically means that given the values for the inputs of a circuit, the values for its outputs are unique. \civer codifies this property as an SMT problem, for which in principle the solver should be able to perform non-linear polynomial reasoning over a prime field. To avoid this, \civer encodes the property using transformation and deduction rules that eliminate finite-field reasoning and then uses \zeethree as the back-end solver to solve the resulting problem in Non-linear Integer Arithmetic (NIA).

 We have instrumented \civer to generate weak-safety encodings for \cvc, \yices and \ffsol  and then we have used it to verify weak safety in the circuits of the circomlib \cite{circomlib}, the standard library of circuits for the circom language. As a result, we have obtained 719 new benchmarks that allow us to perform an experimental comparison among \cvc, \zeethree, \yices and our solver.  In this context, a \sat result means that the weak safety property is not satisfied by the circuit for arbitrary inputs.

\begin{center}
	\begin{minipage}{0.3\textwidth}
		\centering
		\begin{tabular}{c|ccc}
		 \diagbox{{\cvc}}{{\ffsol}} & {\sat} & {\timeout} & {\unsat} \\
			\hline
			{\sat}     & 2 & 2  & -   \\
			{\timeout} & 1  & 5  & 49 \\
			{\unsat}   & -   & 0  & 660 \\
		\end{tabular}
		\captionof{table}{\ffsol vs. \cvc}
		\label{tab:ffsolvscvc5}
	\end{minipage}
	\hfill
	\begin{minipage}{0.3\textwidth}
		\centering
		\begin{tabular}{c|ccc}
		 \diagbox{{\zeethree}}{{\ffsol}} & {\sat} & {\timeout} & {\unsat} \\
			\hline
			{\sat}     & 3 & 1  & -   \\
			{\timeout} & - & 5  & 8 \\
			{\unsat}   & -   & 1  & 701 \\
		\end{tabular}
		\captionof{table}{\ffsol vs. \zeethree}
		\label{tab:ffsolvsz3}
	\end{minipage}
	\hfill
\begin{minipage}{0.3\textwidth}
	\centering
	\begin{tabular}{c|ccc}
		\diagbox{{\yices}}{{\ffsol}} & {\sat} & {\timeout} & {\unsat} \\
		\hline
			{\sat}     & 3 & 2  & -   \\
{\timeout} & - & 5  & 319 \\
{\unsat}   & -   & -  & 390 \\
	\end{tabular}
	\captionof{table}{\ffsol vs. \yices}
	\label{tab:ffsolvsyices}
\end{minipage}
\end{center}

In Table \ref{tab:ffsolvscvc5}, we observe that \cvc performs slightly better on the \sat benchmarks. However, for the \unsat benchmarks, \ffsol solves 49 more instances than \cvc, and moreover does not miss any \unsat benchmark that \cvc is able to solve. Besides, \ffsol proves the 99\% of these instances in an average time of 1.26 seconds, whereas \cvc proves the 92.4\% of them in 3.59 seconds.

As can be seen in Table \ref{tab:ffsolvsz3}, the performance of \ffsol and \zeethree is very similar, with \ffsol also getting slightly better results. However, it is important to highlight that while both \cvc and \ffsol solve exactly the same problem, \zeethree is fed with the NIA problem that is obtained after applying the transformation and deduction rules from \cite{DBLP:conf/sp/IsabelRR24}. In this case, \zeethree proves the 98.2\% of the instances in 0.92 seconds. Thus, our solver performs similarly to \zeethree, but
using a more generic encoding that does not require the specific transformations applied in \civer when proving determinism of constraint systems.
Finally, Table \ref{tab:ffsolvsyices} shows that \yices is the only solver that successfully identifies the five \sat problems, although it only solves 54.9\% of the total instances.

\subsection{Ablation Study of the Modules}

In order to evaluate the contributions of the different modules included in the solver, we performed an ablation study in which we incrementally enabled the modules described in Section \ref{sec:theory-solver}. Tables \ref{fig:ablation1} and \ref{fig:ablation2} show the results on the benchmarks introduced in Sections \ref{exp:cvc5} and \ref{exp:civer}, respectively. Each column specifies the set of enabled modules: C3.$N$ denotes a configuration in which the solver applies the modules described in Sections from $3.1$ to 3.$N$. The results show that all modules have a positive impact on performance. For instance, enabling the linear clause inference module (C3.5) allows the earlier prime field linear module and integer linear module to solve more instances by feeding them with new linear clauses. Overall, the configuration that enables all modules achieves the best behaviour.

\begin{center}
	\begin{minipage}{0.45\textwidth}
		\centering

	\begin{tabular}{c| c c c c c c} 
	& C3.1 &{C3.2} & {C3.3} & {C3.4} & {C3.5} & C3.6 \\
	\hline
	{\sat}  & 0 & 341 & 342  & 342 & 444 & 811 \\
	{\unsat}   & 212 & 411 & 576  & 570 & 669 & 669 \\
	\end{tabular} 
	\captionof{table}{Comparison using \ref{exp:cvc5}}
	\label{fig:ablation1}

	\end{minipage}
	\hfill
	\begin{minipage}{0.45\textwidth}
	\centering
	
	\begin{tabular}{c| c c c c c c} 
	& C3.1	&{C3.2} & {C3.3} & {C3.4} & {C3.5} & C3.6 \\
		\hline
		{\sat}   & 0 & 1 & 1  & 1 & 3 & 3 \\
		{\unsat}  &   309 & 684 & 696  & 705 & 709 & 709 \\
	\end{tabular} 
	\captionof{table}{Comparison using \ref{exp:civer}}
	\label{fig:ablation2}
	
\end{minipage}
\end{center}

\section{Related Work}
\label{sec:related-work}

The literature on satisfiability modulo arithmetic theories is extensive. There is a wide range of fragments that have been considered so far, from difference logic \cite{DBLP:conf/cav/NieuwenhuisO05} and linear constraints \cite{DBLP:conf/cav/DutertreM06}, in particular over integer variables \cite{DBLP:journals/jsat/Griggio12,DBLP:journals/jar/JovanovicM13}, to integer \cite{DBLP:journals/tocl/BorrallerasLROR19} and real \cite{DBLP:journals/jsat/FranzleHTRS07,DBLP:journals/cca/JovanovicM12,DBLP:journals/tocl/CimattiGIRS18,DBLP:journals/jlap/AbrahamDEK21,DBLP:conf/cade/KremerRBT22} non-linear arithmetic. Arithmetic is so ubiquitous in applications that, after more than two decades of the first SMT solvers, research on the design and implementation of efficient arithmetic solvers is still ongoing. For example, in \cite{DBLP:conf/cav/BjornerN24} a new implementation of the arithmetic solver in \zeethree is described. There are some similarities between the method presented in this paper and the module in \zeethree for non-linear real and integer reasoning: in both cases a waterfall approach is applied, and there is a subcomponent for LIA reasoning, another for Gr\"obner basis computation, and another for NRA reasoning. However, these components are combined in a different way due to the underlying variable domains, and most importantly, the key solver for prime field linear arithmetic described in \ref{sec:prime-field-linear-module} is missing in \cite{DBLP:conf/cav/BjornerN24}.

Focusing on SMT on finite fields, a close work to ours is \cite{DBLP:conf/cav/OzdemirKTB23}, where a theory solver for finite fields based on Gr\"obner bases is presented. But while our approach also incorporates Gr\"obner bases, due to their computational cost in our case they are conceived as a last resort to be used only if no other component in the orchestra can make progress. For this same reason, like \cite{DBLP:conf/cav/OzdemirKTB23} we do not introduce field polynomials either, which implies that some unsatisfiable instances may be overlooked and given an \unknown result. On the other hand, we do not try to recover completeness by means of exhaustive enumeration, since in practice the prime numbers of interest are huge (with hundreds of bits). Another difference is that their approach requires to instrument the Gr\"obner basis engine in order to produce explanations for conflicts, while we instead use it as a black box, with the only assumption that the tool is able to express the polynomials in the Gr\"obner basis as a combination of the original generators. This functionality is available in many computer algebra systems \cite{maple,monagan2012maple,DBLP:conf/issac/BigattiR06,M2,Mathematica}. Moreover, from our point of view our method exploits in a better way the benefits of a DPLL($T$) architecture: while in \cite{DBLP:conf/cav/OzdemirKTB23} the theory solver acts only once a full propositional assignment has been constructed and produces no lemmas or theory propagations, our approach makes lightweight checks that test if the partial propositional assignment is consistent with the theory, and if a conflict is detected a lemma is produced and submitted to the SAT solver. Moreover, our theory solver may propagate theory literals as well as generate dynamically clauses implied by the theory. Finally, their approach can be applied to general finite fields, while here we focus on prime fields as this allows us to identify cases in which integer and modular reasoning are equivalent. Nevertheless, if no integer reasoning is to be exploited, our techniques can also be implemented for non-prime finite fields in a straightforward way.

A refinement of the approach in \cite{DBLP:conf/cav/OzdemirKTB23} for instances with 0-1 variables and bitsums is presented in \cite{DBLP:conf/cav/OzdemirPBFBD24}. The notion of split Gr\"obner bases is introduced, which attempts to alleviate the burden of computing full Gr\"obner bases by dividing the problem into smaller subpieces, namely a \emph{linear} module and a \emph{sparse} module. On the other hand, for this kind of instances we propose a component that identifies when modular reasoning reduces to integer reasoning, that is, when variables may be assumed to take values in $\ints$ instead of $\ints_{p}$. Also, while both approaches give a special treatment to linear polynomials, ours has the advantage that linear disequations are handled natively by our linear solver, and therefore do not need to be transformed into quadratic constraints with additional variables as dictated by Rabinowitsch trick \cite{Rabinowitsch1930}. Another similar feature is their \textsf{extraProp} subroutine, which performs reasoning based on congruence closure. However, our congruence closure module is not so tailored to bitsums. Finally, again their theory solver \emph{``acts only once a full propositional assignment is available; it constructs no theory lemmas; and it propagates no literals.''}

Another related work can be found in \cite{DBLP:conf/smt/HaderK22,Hader2002MasterThesis,DBLP:conf/lpar/HaderRK23,DBLP:conf/ijcar/HaderKIGK24}. The authors present an approach for SMT on finite fields based on the framework of the Model Constructing Satisfiability (MCSat) calculus \cite{DBLP:conf/vmcai/MouraJ13}, which requires costly calls to quantifier elimination subprocedures.

Finally, yet another close work is \cite{DBLP:conf/sp/IsabelRR24}, which presents a modular approach for verifying safety properties of arithmetic circuits described as polynomial equations over a finite field. As a result of the decomposition of the verification conditions, SMT problems on finite fields are obtained, which the authors propose to \emph{eagerly} encode as problems of SMT in non-linear integer arithmetic by means of transformation and deduction rules. The techniques proposed here are built upon these rules, and allow applying them not only statically but \emph{dynamically} while the search for a solution is being conducted.




\section{Conclusions and Future Work}
\label{sec:conclusions-and-future-work}

We have presented a new theory solver within the DPLL($T$) framework for solving the satisfiability problem modulo the theory of polynomial equations on a prime field. A key feature of the solver is that it consists of several components with different tradeoffs between performance and completeness, ranging from a cheap but imprecise solver of linear constraints in $\primefield$ to more accurate and costly non-linear reasoning modules. These components are orchestrated so that the most efficient ones are called with more priority, but also taking into account that some of them only make sense when a leaf of the search tree has been reached. The proposed techniques have been implemented in a prototype that already shows better results than state-of-the-art tools not only on existing benchmarks~\cite{DBLP:conf/cav/OzdemirKTB23} but also on new benchmarks originated from the verification of arithmetic circuits for zero-knowledge proofs~\cite{DBLP:conf/sp/IsabelRR24}.

For future work we plan to equip our prototype with proof generation. While satisfiability can be easily checked thanks to the generated model, when the input formula is reported to be unsatisfiable the current implementation does not offer any certificate. This may be particularly critical in applications to verification, where a refutation would formally guarantee the absence of errors. The fact that our prototype resorts to external tools for the computation of Gr\"obner bases makes it even more imperative to offer the user the possibility of validating \unsat results with a third-party proof checker. In another line of research,
we think
there is ample room for improvement in the orchestration of the modules of the theory solver, in particular so as to make it adaptive to the characteristics of the instance under consideration.



\begin{credits}

\subsubsection{\discintname}
The authors have no competing interests to declare that are
relevant to the content of this article.
\end{credits}

\bibliographystyle{splncs04}
\bibliography{paper}

\end{document}